\documentclass[%
 reprint,
 amsmath,amssymb,
 aps,
]{revtex4-1}

\usepackage{graphicx}
\usepackage{dcolumn}
\usepackage{bm}
\usepackage{amsmath}
\usepackage{hyperref}
\usepackage[mathlines]{lineno}
\usepackage{esint}

\begin{document}


\title{On the Electromagnetic Nature of Planck Constant}

\author{Paulo Roberto Bueno}
\email{paulo-roberto.bueno@unesp.br}
\affiliation{Institute of Chemistry, Department of Engineering, Physics and Mathematics, S\~ao Paulo State University, Araraquara, S\~ao Paulo State, Brazil\\
}

\date{\today}

\begin{abstract}
It has been demonstrated that a fundamental quantum rate concept with a frequency of $\nu_e = e^2/ g_e h C_q = E_{gs}/h$ ($g_e$ is a degeneracy of 2) can be defined for the dynamics of electrons at the ground-state energy level $E_{gs} = e^2/g_eC_q$. $\nu_e$ is defined as the ratio between half of the reciprocal of the von Klitzing constant, $R_k = h/e^2 \sim$ 25,812 $\Omega$, and the quantum capacitance $C_q = 2 \lambda \varepsilon_0$ of the ground state, where $\lambda = 2\pi r$ refers to the wavelength of the electron in synchrotronic orbital dynamics of radius $r$ and free-space electric $\varepsilon_0$ and magnetic $\mu_0$ permittivities. Using this $\nu_e$ allied to the magnetic rate $\nu_m$ concept, it is possible to resolve Maxwellian expressions for these orbital electrodynamics and conclude that Planck constant can be stated in terms of $e$, $\varepsilon_0$ and $\mu_0$ fundamental constants of electromagnetism, such as $h = (e^2/2\alpha)\sqrt{\mu_0/\varepsilon_0}$, where $\alpha$ is the dimensionless Sommerfeld constant. The ground-state electromagnetic nature of the electron complies with the relativistic massless Dirac Fermionic electrodynamics, as predicted by the Dirac equation, and is a consequence of the electric orbital energy $E_{gs} = e^2/2g_e \lambda \varepsilon_0 = e\varepsilon_i$, where $\varepsilon_i = \oint_{\scriptstyle\partial \Sigma} \textbf{E} \,d\textbf{l}$ is an induced electromotive force, owing to the Faraday induction law, leading to $\tau_C \varepsilon_i$ (where $\tau_C = 1/2g_e\nu_e$). Because $\tau_C \varepsilon_i$ obeys closed periodic dynamics, in agreement with Maxwell’s law, it is equal to the reciprocal of the Josephson constant that defines the magnetic flux quantum $\Phi_B = h/e$ in orbit, complying with the uncertain principle written as $ e \varepsilon_i \tau_C = h$.

\end{abstract}

\pacs{Valid PACS appear here}
\maketitle


\section{\label{sec:introduction}Introduction}

Quantum rate concept $\nu \propto e^2/hC_q = E/h$ was introduced (in the field of physical chemistry) \citep{Bueno-QR-foundation-2020} as the ratio of the conductance quantum~\citep{Imry-Landauer-1999} $G_0 = g_s e^2/h = 1/R_q$ $\sim$ 77.5 $\mu$S~\footnote{Identified simply as the reciprocal of von Klitzing constant, $R_k = h/e^2 \sim$ 25,812 $\Omega$, multiplied by spin degeneracy $g_s$ of the electron.} and the quantum capacitance $C_q$ to define the chemistries referred to as the electron transfer rate constant of electrochemical reactions, which compute the rate of electron transfer between the donor (reducer) and acceptor (oxidant) states.

Using statistical mechanics, the grand canonical ensemble can be invoked to compute the thermal broadening of this rate, defining the energy difference between the orbital states of reactants and products, which can contribute to the most favorable electronic states for the reaction to proceed to form the products. The electrodynamics responsible for accommodating the bonding energy during this chemical reaction can be expressed as $\Delta E = e^2/C_q = \left( k_{B}T \right) \left[f(1-f) \right]^{-1}$, with an associated rate defined as $\nu = e^2/hC_q = \left( k_{B}T/h \right) \left[f(1-f) \right]^{-1}$, where $f = \left[1 + \exp  \left( \Delta E/k_{B}T \right) \right]^{-1}$ is the Fermi-Dirac distribution~\citep{Bueno-QR-foundation-2020}. Considering an adiabatic electron transfer process as a particular electrochemical reaction case, $\Delta E = - e\Delta V$ is proportional to the electrochemical potential difference $\Delta \mu = -e \Delta V$ between the donor and acceptor states governing the dynamics of electrochemical reactions. As a kinetic-limiting case of any chemical reaction that depends on electron coupling between the reactant and products, a rate following $\nu = \Delta E/h$ ultimately controls the kinetics of these reactions.

For instance, in the limit case of the Boltzmann distribution for $f$, where $\exp \left( \Delta E/k_{B}T \right) \gg 1$, the $\nu$ rate can be defined as

\begin{equation}
 \label{eq:Arrhenius-TST}
	\nu = \frac{k_BT}{h}\exp \left( -\frac {\Delta E} {k_{B}T} \right),
\end{equation} 

\noindent which fundamentally governs thermal-activated, Arrhenius type, processes. Therefore, depending on how statistical mechanics is computed to resolve the thermal broadening over $\nu$ and, consequently, the corresponding $\Delta E$, which eventually can take the form of Gibbs free energy as a particular case, $\nu$ can be considered as a first-principle quantum mechanical approach for dealing with chemical kinetics. For instance, the rate constant predicted by transition state theory~\citep{TST-1996} can be deduced from $\nu \propto e^2/hC_q = E/h$, considering a particular statistical mechanics setting over this conceptual definition~\footnote{The fundamental definition of quantum rate is established as the ratio between the reciprocal of the von Klitzing constant and the quantum capacitance} of $\nu$.

Two other uses of this $\nu$ conceptual rate are illustrated below: (subsection~\ref{sec:rate-electrochemistry}) the electron transport comprising junctions within redox molecular switches~\citep{Alarcon-2022, Bueno-QR-foundation-2020} and (subsection~\ref{sec:rate-graphene}), the Dirac point of a capacitive/conductive V-shape of single-layer graphene~\citep{Bueno-QRGraphene-2022, Lopes-Carbon-2021}. 

The importance of $\nu$ for these two situations, in conjunction to the fact that $\nu$ complies with the Dirac electrodynamics theory~\citep{Bueno-QRGraphene-2022}, was the motivation for further (the goal of the present work) defining a $\nu_e$ concept for investigating the electrodynamics at the ground state. The conclusion following from the analysis of this ground state electrodynamics is that the electron motion complies with a massless Dirac particle~\citep{Novoselov-2005}. This conclusion comes from resolving Maxwellian expressions for the electron considering a particular $\nu_e$ as a fundamental constant of the motion. Ultimately, the analysis of $\nu_e$ as a fundamental concept led to the conclusion that (i) Planck constant, as the constitutive physical constant of quantum mechanics, is purely electromagnetic in nature, and (ii) the electrodynamics of the electron at the ground state, such as electromagnetic radiation, is an inductive electromagnetic phenomenon. 

\subsection{\label{sec:rate-electrochemistry} Quantum Rate Concept in Electrochemistry}

In the context of electrochemical reactions, in which electrodynamics is of fundamental importance, the semi-classical Marcus electron transfer theory~\citep{Marcus-1964, Marcus-1993} (a Nobel prize theory in chemistry) cannot only be deduced by considering a thermal broadening of the energy associated with the conceptual electrochemical $\nu$~\citep{Bueno-QR-foundation-2020}, but also it conducts to a Gibbs free energy setting that accounts for the contribution of the reorganization energy of the solvent~\citep{Alarcon-2022} as an additional external potential contribution for the kinetics of electrons transported in electrolytic environments. However, it also can be demonstrated that electrochemistry, as a type of nature's electronics (as it is through electrochemical signal that the electricity flows in `biological circuitries') is, astonishingly, quantized at room temperature, allowing a minimal energy loss during the electron transport~\citep{Sanchez-QR-efficience-2022}.

The latter situation was demonstrated using different redox molecular switches self-assembled over metallic contacts that served as electron reservoirs to test the concept and resolve the nature of the electron transfer rate process in electrochemistry~\citep{Alarcon-2022}. In particular, for this type of electrochemical junction, electron transfer occurs in a diffusionless situation in which $C_q$ can be experimentally investigated, as the only variable determining the magnitude of $\nu$ because $e^2/h$ is a constant. 

In this type of molecular junction, it has been experimentally demonstrated that the charge transfer resistance~\citep{Sanchez-QR-Rct} is equal to the reciprocal of $G_0$ (within experimental error, it was demonstrated to be approximately 12.9 k$\Omega$, which is a theoretical value that complies with $R_q = 1/G_0 = h/g_se^2$) as a limiting value of the conductance in this type of junction~\citep{Sanchez-QR-Rct}. Interestingly, the charge transfer resistance is at the minimum conductance quantum permitted per electron of the junction, despite the non-adiabatic condition stated in this experimental setting~\citep{Sanchez-QR-Rct}.

\subsection{\label{sec:rate-graphene} Quantum Rate at the Dirac Point}

An equivalent $\nu$ concept was defined to investigate the electrodynamics at the Dirac point in single-layer graphene, which corresponds to the minimum of a capacitive or conductive V shape ~\citep{Bueno-QRGraphene-2022, Lopes-Carbon-2021}. It was also demonstrated that $\nu$ established for graphene conforms to the energy-wave vector dispersion ($E-|\textbf{k}|$), as predicted by Dirac electrodynamics theory~\citep{Dirac-1928}, such as $E = \hbar \textbf{c}_* \cdot \textbf{k} = \textbf{p} \cdot \textbf{c}_*$, where $\textbf{k}$ is the wavevector, $|\textbf{k}|$ its modulus, \textbf{c}$_*$ is the vector representation of the Fermi velocity $c_*$, and $\textbf{p} = \hbar \textbf{k}$ is the momentum vector. This linear $E-|\textbf{k}|$ relationship can be verified by observing that $\nu = G_0/C_q = c_*/\lambda$, defining the properties of electrons transported through ideal quantum channels~\footnote{Non-ideal situations, corresponding to a non-adiabatic settings of the electron transmittance, can be treated using transmission probability matrix for accounting to a probability of the transmission of the electron that is lower than the unit.} (that adiabatically connect orbital states during chemical reactions, as stated in sections \ref{sec:introduction} and \ref{sec:rate-electrochemistry}).

The fact that $\nu$ predicts an equivalent $E-|\textbf{k}|$ dispersion to that of the Dirac equation~\citep{Dirac-1928} can be easily demonstrated by noting that $\nu = c_*/\lambda \propto e^2/hC_q \propto E/h$, which can be rearranged as $E = \hbar \textbf{c}_* \cdot \textbf{k} = \textbf{p} \cdot \textbf{c}_*= h\nu$. Accordingly, one of the main conclusions obtained by applying $\nu$ to describe the dynamics of electrons in single-layer graphene is that \textit{at the Dirac point neighborhood, there is relativistic behavior, but outside of the Dirac point region, a non-relativistic treatment is required}. 

This conclusion, attained by employing the concept of $\nu$ in a two-dimensional setting, confirms the importance of a relativistic approach in describing the electrodynamics at the Dirac point for graphene. The explicit treatment using a time-dependent analysis, in which $\tau = 1/\nu$ is a fundamental time/rate concept, permitted the experimental demonstration that \textit{the maximum electron rate dynamics are acquired precisely at the Dirac point}~\citep{Bueno-QRGraphene-2022}, where presumably the net carrier concentration and the DC conductance are null, as observed in a traditional field-effect configuration of measurement. The origin of the maximum electron rate has been demonstrated to be associated with a dynamic displacement current~\citep{Bueno-QRGraphene-2022}. \textit{Particularly in graphene, the relativitic analysis is important specially at the Fermi level, corresponding to a maximum rate $\nu$ at the Dirac point of graphene}~\citep{Bueno-QRGraphene-2022}. For instance, the practical consequence is that electrons in graphene can be transported at a maximum rate at the Dirac point using AC instead of the DC approach for electrical measurements. 

Nonetheless, this particular electrodynamical characteristic of $\nu$ has not only been attained in graphene, but it has also been observed in the transport of electrons in molecular redox switches (subsection~\ref{sec:rate-electrochemistry}), where the net electrochemical current of the junction is also null at the Fermi level, but using an AC mode of perturbation, the maximum displacement electrochemical current (proportional to $C_q$) is measurable.

\section{\label{sec:QRtheory}Electrodynamics At the Ground State Energy Level}

\subsection{\label{sec:Degeneracy} Degeneracy of The Electric Potential Energy}

To begin using the $\nu$ concept for describing the electrodynamics at the ground state, denoted herein as $\nu_e$, for this particular orbital dynamics, it must first be noted that, as in all other previous situations, the formulation of $\nu_e$ departs from the definition of an electrochemical capacitive state denoted as $C_\mu$~\cite{Buttiker-1993, Bueno-book-2018, Iafrates-1995}, which is a series ($1/C_\mu = 1/C_e + 1/C_q$) combination of `electrostatic' $C_e$ and quantum (dynamical) $C_q$ (alternatively referred to as chemical~\cite{Bueno-book-2018, Bueno-capacitance-2019, Bueno-Davis-2020}) capacitive contributions. Nonetheless, it is appropriate to develop an analysis in terms of an electric potential energy representation of this phenomenon, which is equivalent to capacitive depiction, but more convenient for investigating the electrodynamics in the ground state. 

Therefore, it can be noted that a capacitive analysis at the ground state involves only the elementary charge (but with two contributions to the energy state) of the electron in its orbital dynamics, which corresponds to a total electric potential energy of $eV_{\mu}$ that computes the `static' (owing solely to the charge of the electron \textit{per se}) and dynamic (related to the closed motion of the electrons around the nuclei) contributions. Accordingly, at the ground state setting, the total electric potential energy is given by the sum of these two distinct contributions as $eV_{\mu} = eV_e + eV_q$, where $eV_e = -e/4\pi\varepsilon_0r$ and $V_q = -e/2\lambda\varepsilon_0$ are numerically equivalent owing to the mathematical fact that $2\lambda\varepsilon_0 = 4\pi\varepsilon_0r$. 

Nevertheless, the physical origin of each of the above-defined electric potentials (or energies) is distinct, in accordance with the Maxwellian laws of electrodynamics~\citep{ED-Griffiths-1987}, which states that there are two origins for the electric field: one (1) due to the elementary charge of the electron and the other (2) owing to the orbital motion of the electrons, induced by the time variations of the magnetic field. Recall that the time variations in the magnetic field induce an electromotive force $\varepsilon_i$ as predicted by Faraday’s law~\citep{ED-Griffiths-1987}. Accordingly, $eV_e$ is associated with the charge of the electron \textit{per se}, in agreement with the Gauss law, whereas $eV_q = \varepsilon_i$ is due to a different physical nature, purely associated with an inductive electromagnetic phenomenon related to the synchrotronic orbital dynamics in the ground state (more details below). 

The superposition of these two potential energy states (identified in the traditional quantum mechanical analysis as the particle-wave duality of the electron) is well-established and has been formulated in terms of the degeneracy ($g_e = 2$) of the energy state of the electron. This $g_e$ degeneracy is generally associated with the intrinsic spin character of the electron. Hence, to maintain consistency with traditional quantum mechanics, from now on, it will be considered only half of the above-analysed source of electric potential energies (associated to the physical meaning of $e\varepsilon_i = eV_q$ instead of $eV_e$), such as $\nu_e = e^2/g_ehC_q$, where $g_e$ is computed to preserve consistency with the traditional nomenclature. Therefore, the dynamical character of the potential energy associated with the electromotive force $\varepsilon_i$ will be considered in physical analysis below, instead of the traditional `static' component of the electric potential.

In the following section, it will be used the electric $\nu_e = e^2/g_ehC_q$ and an additional magnetic $\nu_m$ quantum rate to describe the closed orbital-like motion of the electron in its ground state energy level. Using these rates to compute the periodic oscillatory motion of an electron, it is possible to comply with the Faraday and Maxwell inductive laws of electrodynamics~\citep{ED-Griffiths-1987} and demonstrate that the Planck constant $h$ is essentially electromagnetic in nature.

The electromagnetic nature of $h$ will be the basis for further analysis~\citep{Bueno-2023} of this quantum electromagnetic approach, where it will be demonstrated that the dimensionless Sommerfeld (or fine-structure) $\alpha$ constant~\citep{Sommerfeld-1916} can be expressed solely in terms of electric $1/\tau_C = \left( g_sg_e \right) \nu_e$ and magnetic $1/\tau_L = \nu_m$ rates in a coherent quantum electromagnetic approach to the electrodynamics of the ground state.

\subsection{\label{sec:Maxwell} Maxwellian Equations for the Electrodynamics of the Ground State}

Gauss’s law for magnetism~\citep{ED-Griffiths-1987}, in the setting of the motion of an electron in a closed orbital trajectory, is formulated as $\oiint_{\partial\,\mho}\textbf{B}\;d\textbf{S} = 0$, whereas the elementary charge $e$ contributes to the electric potential as $\Phi_E = \oiint_{\partial\,\mho}\textbf{E}\;d\textbf{S} = -e/\varepsilon_0$. For both the above-described magnetic and electric formulations of Gauss’s law, $\mho$ states for any orbital volume with a closed boundary surface $\partial\,\mho$. Faraday and Maxwell’s laws are stated for the closed orbital dynamics of the electron within $1/\tau_C$ and $1/\tau_L$ rates, respectively, as

\begin{equation}
 \label{eq:Faraday-law}
	\oint_{\scriptstyle\partial\,\Sigma} \textbf{E} \,d\textbf{l} = - 2\pi \frac{\Phi_B}{\tau_C} = |\textbf{E}| \lambda,
\end{equation} 

\noindent and

\begin{equation}
 \label{eq:Maxwell-law}
	\oint_{\scriptstyle\partial\,\Sigma} \textbf{B} \,d\textbf{l} = \mu_0 \varepsilon_0 \frac{\Phi_E}{\tau_L} = |\textbf{B}| \lambda,
\end{equation}

\noindent where $\Phi_B = h/e = \iint_{\partial\,\Sigma}\textbf{B}\;d\textbf{S}$ in Eq.~\ref{eq:Faraday-law} is the magnetic flux quantum and $\tau_L$ is the periodic oscillatory characteristic time of the orbital magnetic dynamics, as in the case of $\left( g_s g_e \right) \tau_C = 1 / \nu_e$ for the electric contribution. $|\textbf{E}|$ and $|\textbf{B}|$ are the magnitudes of the electric $\textbf{E}$ and magnetic $\textbf{B}$ field vectors~\footnote{Although it is not usual, it was adopted this notation to avoid misinterpretation of the magnitude of electric field vector $|\textbf{E}|$ with the notation assumed for the energy $E$.}. Observe that in Eq.~\ref{eq:Faraday-law} and Eq.~\ref{eq:Maxwell-law}, $|\textbf{E}| \lambda$ and $|\textbf{B}| \lambda$ are the results of the integrand operation over any orbital surface $\Sigma$ with a closed boundary curve $\delta \Sigma$. Finally, it must be observed that the Ampère circuital law contribution to Eq.~\ref{eq:Maxwell-law} was disregarded owing to the absence of a continuous flow of electric current~\footnote{In other words, there is not a direct current (DC) mode of electric current flowing in this type of oscillatory motion.} in this type of wave dynamics.

\subsubsection{\label{sec:Faraday-law} The Relativistic Interpretation of Faraday Law For the Electrodynamics of Ground State Energy Levels}

Observe that according to Eq.~\ref{eq:Faraday-law}, the electromotive force $\varepsilon_i$ induced by the magnetic flux quantum can be expressed as follows: 

\begin{equation}
 \label{eq:Faraday-law-alternative}
	\varepsilon_i = - \omega_C \Phi_B,
\end{equation}

\noindent where $\omega_C = 2 \pi / \tau_C = \left( g_s g_e \right) \omega_e$. For example, substituting $\Phi_B = \hbar / e$ and $\omega_e = g_e \hbar C_q/e^2$ in Eq.~\ref{eq:Faraday-law-alternative} leads to the conclusion that $\varepsilon_i = -e/g_e C_q$. Alternatively, using the modulus of $\varepsilon_i$ and noting that the energy is $e \epsilon_e =  h\nu_e = \hbar \omega_e$, with further basic algebraic operations, it is possible to demonstrate that $h\nu_e = e^2/g_e C_q = e^2 / 2 g_e \lambda\varepsilon_0$. This physical analysis is consistent with the interpretation of traditional quantum mechanics, as expected. For instance, using the Bohr radius $r_b \sim 5.29 \times 10^{-2}$ nm as a reference for the orbital dynamics with a wavelength of $\lambda = 2\pi r_b$, it leads to $h\nu_e = e^2/2 g_e \lambda\varepsilon_0 \sim 13.6$ eV as the energy of the ground state, which is in agreement with the experimentally observed value for the hydrogen atom. 

Furthermore, this physical analysis associated with an induced electromotive energy $e\varepsilon_0$ interpretation of the quantum mechanics of the ground state is consistent with Dirac electrodynamics~\citep{Dirac-1928}, permitting a straightforward interpretation of the massless relativistic character of electrodynamics in the ground state, as will be demonstrated here. 

It must be noted that Eq.~\ref{eq:Faraday-law-alternative} is also consistent with the discussion conducted in Section \ref{sec:rate-graphene}, where it was noted that Dirac electrodynamics~\citep{Dirac-1928} generally comply with a rate obeying $\nu = c_*/\lambda$. Hence, the linear energy-wave vector ($E-|\textbf{k}|$) dispersion relationship suitably describes the electrodynamics in graphene~\cite{Novoselov-2005}. Therefore, applying a similar analysis to the dynamics of the electrons in orbital motion and noting that $\Phi_B = h/e$, $\omega_C = \left( g_sg_e \right) \omega_e$ and $\nu_e = c_* / \lambda$, Eq.~\ref{eq:Faraday-law-alternative} can be rewritten as $e \varepsilon_i = h \nu_e = h c_* / \lambda$: Now, using De Broglie's hypothesis $h = p \lambda$ and rearranging $e\varepsilon_i = h c_*/\lambda$, we obtain: 

\begin{equation}
 \label{eq:E-relativistic}
	E_{gs} = \hbar \omega_e = \textbf{p} \cdot \textbf{c}_* = \hbar \textbf{c}_* \cdot \textbf{k} = -e\oint_{\scriptstyle\partial\,\Sigma} \textbf{E} \,d\textbf{l},
\end{equation}

\noindent which, finally, can be simplified as

\begin{equation}
 \label{eq:E-closed-work-relativistic}
	E_{gs} = \textbf{p} \cdot \textbf{c}_* = -e\oint_{\scriptstyle\partial\,\Sigma} \textbf{E} \,d\textbf{l} = \oint_{\scriptstyle\partial\,\Sigma} \textbf{F}_i \,d\textbf{l},
\end{equation}

\noindent in which $\textbf{F}_i = -e\textbf{E}$ is an electric force associated with the electron orbital dynamics, performing a work $\oint_{\scriptstyle\partial\,\Sigma} \textbf{F}_i \,d\textbf{l}$ over a closed boundary curve trajectory $\partial\,\Sigma$, which is ultimately responsible for the electromagnetic wave character of the electron, such as $\lambda = \oint_{\scriptstyle\partial\,\Sigma} \,d{l}$, arising from the elementary charge dynamics over its periodic oscillatory orbital trajectory in the atom. It is implicit that the relativistic character of this closed work is induced by $\varepsilon_i$, as defined in accordance with Faraday’s law, and by Dirac relativistic theory~\citep{Dirac-1928} owing to its equivalence with $\textbf{p} \cdot \textbf{c}_* = \hbar \textbf{c}_* \cdot \textbf{k}$. 

Furthermore, according to the theory of relativity, a massless relativistic character at the ground state is observed owing to $E_{gs} = \sqrt{\left( \textbf{p} \cdot \textbf{c}_* \right)^2 + m_0^2c_*^4}$, where $m_0$ is the resting mass of the electron, which is assumed to be null for the relativistic case, which is precisely the setting established for the ground state orbital closed trajectory (not dependent on the resting mass, but solely of its electromagnetic inductive nature owing to $e\varepsilon_i = eV_q$); thus leading to $E_{gs} = \textbf{p} \cdot \textbf{c}_* = \oint_{\scriptstyle\partial\,\Sigma} \textbf{F}_i \,d\textbf{l}$ and conforming with an induced work interpretation of the phenomenon owing to the closed oscillatory trajectory of the electron in the ground state.

In the next section, it will be demonstrated how this physical interpretation of the electrodynamics of the ground state implies a reinterpretation of Planck constant.

\subsubsection{\label{sec:Faraday-law} The Electromagnetic Interpretation of Planck Constant}

Eq.~\ref{eq:Maxwell-law} can be rewritten in a simpler form as $|\textbf{B}| \lambda c^2 = \Phi_E/\tau_L$ by noting that $c = 1/\sqrt{\mu_0\varepsilon_0}$ and further simplified as $c^2 = \left( 2r/\tau_L \right) c_*$ by observing that $\Phi_E = 2 \lambda r |\textbf{E}| = 2 \lambda \oint_{\scriptstyle\partial\,\Sigma} \textbf{F}_i \,d\textbf{l}$~\footnote{Note that by doing this procedure it implies the combination of equations \ref{eq:Faraday-law} and \ref{eq:Maxwell-law}.} and $c_* = |\textbf{E}|/|\textbf{B}|$. As the fine structure (or Sommerfeld) constant for ground state orbital electrodynamics is defined as $\alpha = c_*/c \sim 1/137$, the result can be rearranged as $\alpha = \left( c/2r \right) \tau_L$ or

\begin{equation}
 \label{eq:alpha}
	\alpha = \frac{c\mu_0}{4R_q} = \frac{1}{4R_q}\sqrt{\frac{\mu_0}{\varepsilon_0}}= \frac{Z_0}{4R_q},
\end{equation}

\noindent by assuming a definition of $\tau_L = L_q/R_q$ for the magnetic rate of the electron in the orbit, where $L_q = \mu_0 r/2 = \mu_0 \lambda/4\pi$. 

In Eq.~\ref{eq:alpha}, it can be seen that $c\mu_0 = \sqrt{\mu_0/\varepsilon_0} = Z_0$ is the characteristic impedance of free space ($\sim$ 377 $\Omega$). Considering the definition of $G_0 = 1/R_q = g_s e^2/h$ in Eq.~\ref{eq:alpha} and noting that $g_s$ equals to 2, Planck constant $h$ is defined as follows:

\begin{equation}
 \label{eq:h}
	h = \frac{e^2}{2\alpha}\sqrt{\frac{\mu_0}{\varepsilon_0}},
\end{equation}

\noindent which establishes an electromagnetic meaning for $h$ within the above-introduced quantum electromagnetic approach of the ground state dynamics and consequently to quantum mechanics owing to $\alpha$ be a dimensionless fundamental constant that quantifies the strength of electromagnetic interaction between elementary charged particles~\cite{MacGregor-2007}. 

The supposition of a magnetic rate $\nu_m = 1/\tau_L = R_q/L_q$ to describe the magnetic dynamics of the ground state requires separate and dedicated further work owing to the fact that $\alpha$ is related to electromagnetic coherent phase dynamics, as introduced elsewhere~\citep{Bueno-2023}, and defined, in a quantum electromagnetic rate approach, in terms of the ratio between $1/\tau_C$ and $1/\tau_L$ quantum rates, a demonstration that is beyond the scope of this study.

\section{\label{sec:conclusion}Conclusion}

It was demonstrated that there is an additional origin of the electric potential energy of an electron orbiting a nuclei that can be, alternatively to that associated to the elementary `static' charge of the electron as introduced by Bohr, formulated in terms of an electromotive force associated with the closed motion of the electron around the nuclei. This permitted the resolution of the Maxwellian laws of classical electrodynamics within electric $1/\tau_C$ and magnetic $1/\tau_L$ quantum rate settings for describing the time-dependent oscillatory dynamics of the electron in the ground state, conducting to a reinterpretation of the meaning of $h$ in quantum mechanics, as stated by Eq.~\ref{eq:h}.

Notably, the quantum electromagnetic rate reinterpretation of the ground state electrodynamics, as introduced in this work, is not only in compliance with Maxwellian laws~\citep{ED-Griffiths-1987}, but also conforms with the relativistic quantum electrodynamics as proposed by Dirac~\citep{Dirac-1928}; hence, it exhibits an inherent massless character of the electron~\citep{Novoselov-2005} that fulfils both relativistic quantum dynamics~\citep{Dirac-1928} and Maxwellian laws of classical electrodynamics~\citep{ED-Griffiths-1987}.

\section{\label{sec:Acknowledgement}Acknowledgement}

The author acknowledges the Brazilian National Research Council and São Paulo Research Foundation (2017/24839-0).

\section{\label{sec:Data}Data Availability}

All data generated or analysed during this study are included in this published article.

\bibliography{references}
\end{document}